\documentclass[aps,prl,preprint]{revtex4-1}
\usepackage{graphicx}
\usepackage{bm}
\begin{document}
\title{High-frequency acoustic wave detection in Schottky diodes:
theory consideration.}
\author{B.~A.~Glavin}
\address{V.~E.~Lashkaryov Institute of Semiconductor Physics,
Pr.~Nauki 41, Kyiv 03028, Ukraine}

\begin{abstract}
In this paper the theory of high-frequency acoustic signal detection by Schottky diodes is presented. Physically, the detection was found to be due to the quasi-static screening of the potential perturbation caused by the acoustic strain by charge carriers. The total charge required for screening changes with the value of strain at the edge of the semiconductor depletion region and metal-semiconductor interface
giving rise to displacement current. 
The magnitude and frequency dependence of the electrical signals are analyzed for both piezoelectric and deformation potential coupling mechanisms. The obtained results are in good agreement with the recent experimental observations and suggest feasibility of high-frequency (up to terahertz band) acoustic wave detection provided that proper electrical measuring scheme is available.
\end{abstract}


\maketitle

\section{Introduction}

In recent decades acoustic techniques in solid state physics demonstrate serious progress, especially by moving to previously unattainable high-frequency band, up to a terahertz frequencies \cite{ps-ultrasonics}.
Considerable efforts in this direction, called often picosecond ultrasonics, are stimulated by the short-wavelength character of acoustic 
waves in this band, and, in some cases, efficient coupling of acoustic strain to electronic, optical, magnetic 
excitations in solid state. This allows application of high-frequency acoustic signals for testing and control of nanodimensional solid-state structures.  
From practical point of view, the most serious restriction of picosecond ultrasonics is the use of ultrafast (femtosecond) lasers for 
both excitation of acoustic signals and detection of its coupling to a solid-state nanostructure, usually with the use of 
pump-probe technique. In spite of considerable improvement of such lasers characteristics and 
growth of their availability, development of a robust electrically controlled 
picosecond acoustic technique would be an essential breakthrough in the field. Speaking about the high-frequency acoustic wave excitation, 
terahertz sasers could be a solution of the problem \cite{saser1,saser2,saser3}. For detection purposes, several options are available. The superconductor based 
detectors are in use from 70th. The robust bolometers used to be widely employed for acoustic spectroscopy are currently less popular since,
in contrast to optical methods, they are hardly sensitive to the spectrum of an acoustic signal. The superconductor contacts do possess spectral selectivity
\cite{super-contacts}   but their fabrication is quite sophisticated. Semiconductor-based   approaches are more preferable. A photo-electric 
acoustic wave detection by {\it p-i-n} diodes with a quantum well embedded into the {\it i} region demonstrated high efficiency, but, although 
based on electric current measurement, requires use of femtosecond laser for temporal signal sampling \cite{pin}. 
An alternative, also semiconductor-based, method using Schottky diodes has been demonstrated recently \cite{schottky}. It is purely electrical 
and is based on induction of displacement current by propagating acoustic wave. 
Considering such factors as all-electrical detection principle, use of robust well-studied devices technology which can be 
integrated with various solid-state structures, possible room-temperature applications, this method looks an attractive candidate for wide use as 
high-frequency acoustic detector.  In this paper the 
main physical principles of Schottky diode acoustic detection are considered theoretically in details. The developed 
model allows to address such issues as feasible magnitude of the electrical signal, fundamental restrictions on detectable acoustic signal frequency,
possible ways of the diode structure optimization. The paper is organized as follows. In section I the expression for the accumulated electrical charge due to 
the acoustic strain perturbation is obtained for important cases of piezoelectric and   deformation potential coupling. It is used then in section II for analysis of the 
electrical response of the Schottky diode. Then, the conclusions follow.     

\section{Expression for the acoustic wave induced charge in a diode}
\begin{figure}
\centering
 \includegraphics[width=0.6\linewidth]{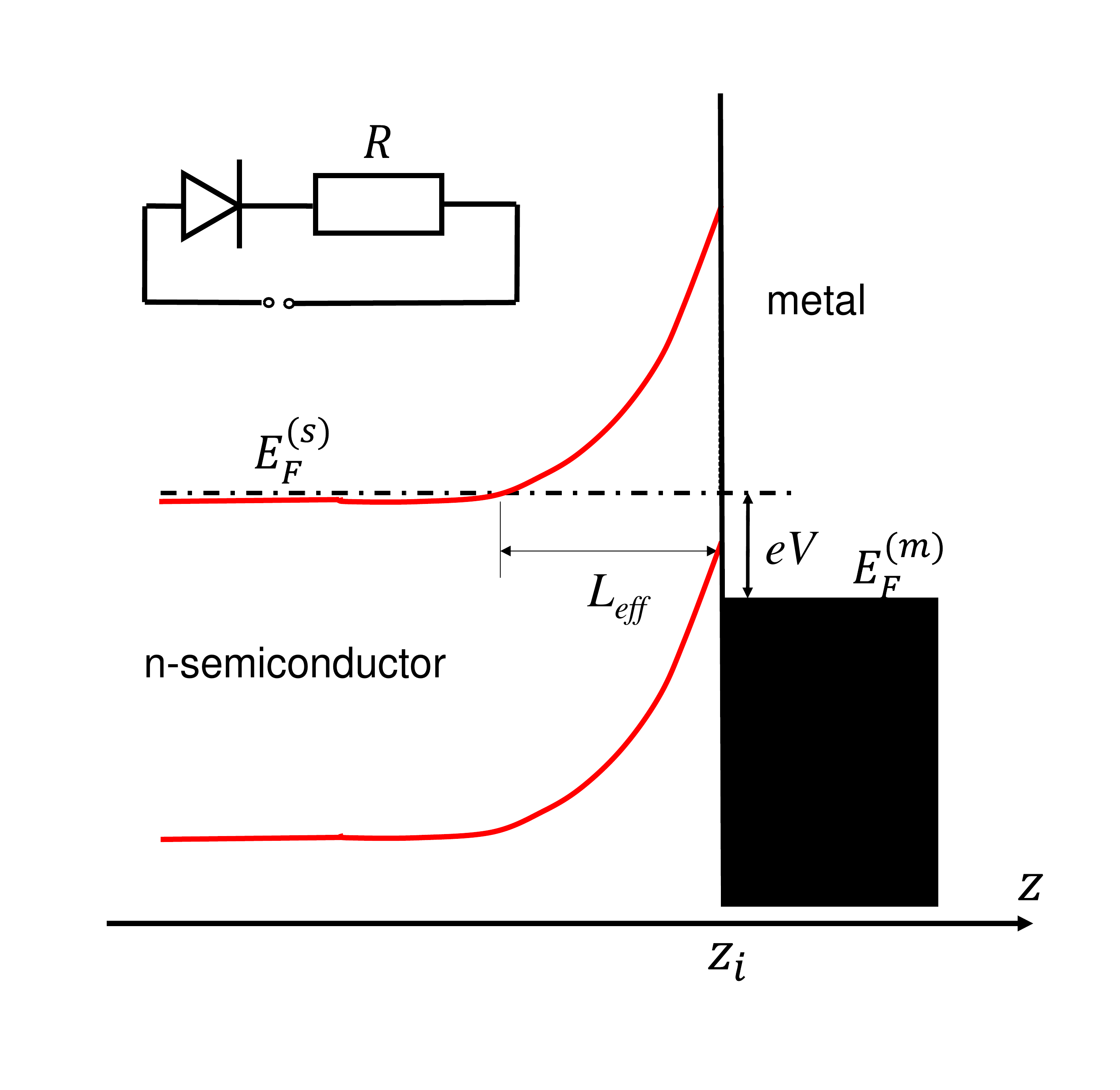}
\caption{The schematics of the energy diagram of $n$-type Schottky diode with the used coordinate frame. 
$z=z_i$ corresponds to the metal-semiconductor interface. The insert shows the model 
electrical circuit which is used for the electrical detection of acoustic signals. 
}
\label{fig:1}
\end{figure}

The energy diagram of the Schottky diode is shown in Fig.\ref{fig:1} for particular case of n-doped semiconductor. 
We consider the range of external biases $V$, for which the Schottky barrier is much higher than temperature measured in 
energy units. In this case the electrical current is small and it is possible to assume that electron distributions in 
semiconductor and metal regions correspond to quasi-equilibrium and can be characterized by quasi-Fermi levels shifted by the value of electrical bias $eV$ assuming positive sign for 
the direct bias of the diode. 
While an acoustic wave propagates through the structure, the related strain induces the potential acting on electrons. Such a potential can be 
described within the deformation potential model \cite{Gantmakher-Levinson}. Redistribution of the charge carriers in this potential gives rise 
to the perturbation of the 
electric filed in the system. In addition, in a piezoelectric semiconductor electric field is perturbed due to the lattice polarization induced by the
acoustic wave. The perturbed potential $\delta \varphi$ satisfies the Poisson equation, 
which in one-dimensional limit, corresponding to an acoustic wave propagating along
$z$-axis which is normal to the
flat metal-semiconductor interface, is 
\begin{equation}
\label{eq:Poisson}
\frac{d \delta \varphi}{dz} =\frac{e}{\varepsilon_s \varepsilon_0} \delta n +
\frac{1}{\varepsilon_s \varepsilon_0} \frac{d P_z}{dz},
\end{equation}
where $\delta n$ is perturbation of the electron concentration, $\varepsilon_s$ and $\varepsilon_0$ 
are the dielectric constant of the semiconductor and the absolute permittivity, and $P_z$ is the 
$z$-component of the peizoelectric polarization. 
The important assumption we are going to use is that all perturbations caused by the acoustic wave are 
much slower than the electron relaxation processes in both metal and semiconductor. The latter 
can be characterized by the dielectric relaxation time $\varepsilon_{s,m} \varepsilon_0/\sigma$, where 
$\sigma$ is conductivity and $\varepsilon_m$ is the lattice dielectric
permittivity of metal. Such time is usually within subpicosecond band for semiconductor 
and even shorter for metal. Thus, we may use a 
quasi-static approach determining $\delta n$ while dealing with sub-terahertz acoustic waves. 
This means that at any time instant the electron density
perturbation is the same as in the case of static nonuniform strain distribution
corresponding to this particular time. Specifically, dropping the time dependence for brevity, in the linear approach for
semiconductor region $z<z_i$ we have:
\begin{equation}
\label{eq:dn_s}
\delta n (z) = e (\delta \varphi (z) - U_{DP}(z)/e - \delta V_s) \frac{dn_s}{dE_F},
\end{equation}
where $n_s (E_F)$ is  the electron concentration dependence on the Fermi energy,
$U_{DP}$ is the deformation potential energy of electrons, and we allow 
perturbation of the semiconductor reference potential, $\delta V_s$, caused by the acoustic wave. 
Note, that the value of 
the derivative in right hand side of Eq.(\ref{eq:dn_s}) depends on coordinate.   
Analogously, in metal, $z>z_i$, we have 
\begin{equation}
\label{eq:dn_m}
\delta n (z) = e (\delta \varphi (z) - U_{DP}(z)/e - \delta V_m) \frac{dn_m}{dE_F}.
\end{equation}
With Eqs.(\ref{eq:dn_s},\ref{eq:dn_m}), the Poisson equation becomes a linear inhomogeneous  
differential equation. 
It is convenient to perform its solution separately for $z<z_i$ and $z>z_i$, 
applying then the boundary conditions at $z=z_i$. Using standard 
variation of constants 
method and taking into account that $\varphi(z=-\infty) = \delta V_s$,
$\varphi(z=\infty) = \delta V_m$ , we obtain 
\begin{eqnarray}
\label{eq:pt-sol}
\delta \varphi (z) = \delta V_s +c_s \phi_{s2} (z) +\frac{\phi_{s2} (z)}{w_s} \int_z^{z_i} dz' \phi_{s1}(z') \left(k_s^2 (z') U_{DP}(z')/e-
\frac{1}{\varepsilon\varepsilon_0} \frac{d P_z}{dz'}\right)  + \\
\frac{\phi_{s1} (z)}{w_s} \int_{-\infty}^{z} dz' \phi_{s2}(z') \left(k_s^2 (z') U_{DP}(z')/e-
\frac{1}{\varepsilon\varepsilon_0} \frac{d P_z}{dz'}\right), \mbox{~for~} z<z_i \nonumber \\ 
\delta \varphi (z) = \delta V_m +c_m \phi_{s2} (z) -\frac{\phi_{m2} (z)}{w_m} \int_{z_i}^z dz' \phi_{m1}(z') k_m^2 (z') U_{DP}(z')/e 
- \nonumber \\
\frac{\phi_{m1} (z)}{w_m} \int_z^{\infty} dz' \phi_{m2}(z') k_m^2 (z') U_{DP}(z')/e, \mbox{~for~} z>z_i. \nonumber
\end{eqnarray}
Here $c_s$ and $c_m$ are constants, $k_{s,m}^2 =e^2dn_{s,m}/dE_{F} (\varepsilon_{s,m} \varepsilon_0)^{-1}$, 
and $\phi_{m1,2}$ are fundamental solutions of the homogeneous versions of 
equations for $\delta \varphi$:
\begin{eqnarray}
\label{eq:pt-hom}
\frac{d \phi_{s1,2}}{dz^2} =k_s^2 (z) \phi_{s1,2} \mbox{~for~} z<z_i \\
\frac{d \phi_{m1,2}}{dz^2} =k_m^2 (z) \phi_{m1,2} \mbox{~for~} z>z_i \nonumber
\end{eqnarray}
These functions are selected such that $\phi_{s2} (-\infty) =0$, $\phi_{m2} (\infty) =0$ 
and Wroskians in Eq.(\ref{eq:pt-sol}) are 
$w_{s,m} = \phi_{s,m1} \phi'_{s,m2} - \phi'_{s,m1} \phi_{s,m2}$.

The constants $c_s$ and $c_m$ are determined via the boundary conditions at $z=z_i$, requiring continuity of potential and electrical 
induction. Then, it is straightforward to calculate the perturbation of the accumulated charge, $\delta Q$:
\begin{equation}
\label{eq:charge-def}
\delta Q=\varepsilon\varepsilon_0 S \int_{z_i}^{\infty} dz k_m^2 (-\delta \varphi (z) + U_{DP}(z)/e + \delta V_m),
\end{equation}
where $S$ is the diode cross-section. After some algebra from the expressions for the potential we obtain
\begin{eqnarray}
\label{eq:charge-expr}
\delta Q=C \left( \delta V - V_{PZ}(z_i) + \int_{-\infty}^{z_i} dz G_s (z) \left( V_{DP}(z) +V_{PZ} (z)\right) - \right. \nonumber \\
\left.  \int_{z_i}^\infty dz G_m (z) V_{DP} (z) \right),
\end{eqnarray}
where we introduced the effective potential due to the deformation potential acousto-electric coupling 
$V_{DP} \equiv - U_{DP}/e$, potential induced due to poiezoelectric action of the aoustic wave 
$V_{PZ}$ such that $V'_{PZ} =   P_z/(\varepsilon_s \varepsilon_0)$, the kernel functions
\begin{eqnarray}
\label{eq:kernel}
G_s (z)= \frac{1}{\phi'_{s2}(z_i)}\phi_{s2}(z)k_s^2(z), \\
G_m (z)= 
\frac{1}{w_m} \left( \phi_{m1}(z_i) \phi'_{m2} (z_i)-\frac{\varepsilon_s}{\varepsilon_m}\phi_{m2} (z_i) \phi'_{m1}\right)
\frac{1}{\phi'_{m2}(z_i)}
\phi_{m2}(z)k_m^2(z) \nonumber
\end{eqnarray}
and the diode capacitance $C=\varepsilon_s\varepsilon_0 S/L_{eff}$ with 
\begin{equation}
\label{eq:thickness}
L_{eff}=\frac{\phi_{s2}(z_i)}{\phi'_{s2}(z_i)} - \frac{\varepsilon_s}{\varepsilon_m} 
\frac{\phi_{m2}(z_i)}{\phi'_{m2}(z_i)}.
\end{equation}

\begin{figure}
\centering
    \includegraphics[width=0.7\linewidth]{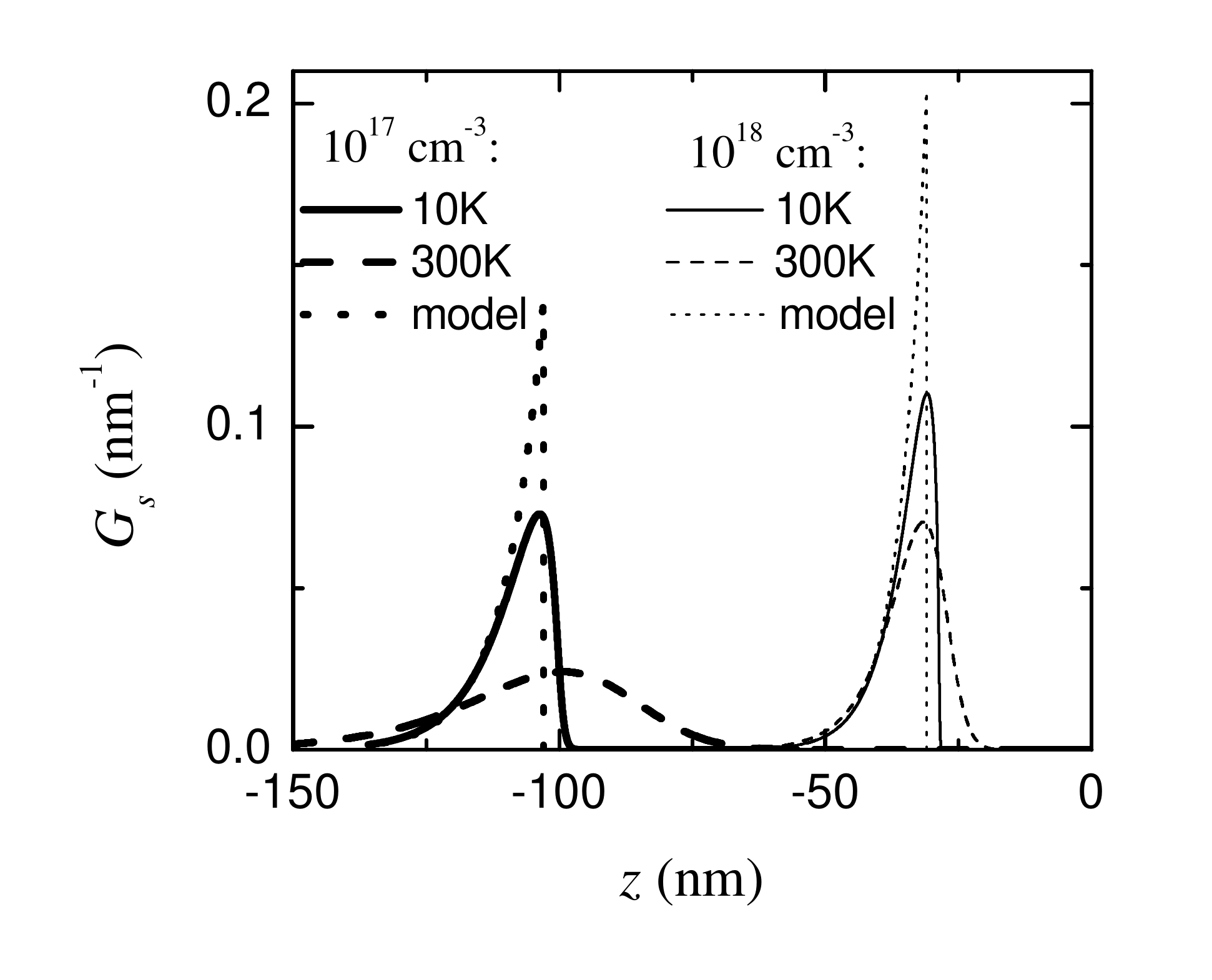}
\caption{The kernel function $G_s$ for GaAs Schottky diodes with doping level $10^{17}$~cm$^{-3}$ and 
$10^{18}$~cm$^{-3}$ and different temperatures. $z=0$ corresponds to the metal-GaAs interface, {\it i.e.}
$z_i=0$. For comparison, the results for model step-like spatial dependence of $k_s^2$ are shown. 
}
\label{fig:kernel}
\end{figure}

In Fig.\ref{fig:kernel} we plot the spatial dependence of the kernel function 
$G_s$ calculated for GaAs Schottky diodes with doping $10^{17}~cm^{-3}$ and 
$10^{18}~cm^{-3}$ and temperatures $10K$ and $300K$. The steady-state potential profile 
and the screening parameter  were determined with the standard approach assuming low
value of the diode current \cite{Sze}. As we see, the charge is controlled by the
perturbation near the edge of the depletion layer. This result is expectable: indeed,
the used boundary conditions assume no acoustic perturbation for $z=-\infty$. In this case 
although variation of strain inside the spatially uniform portion of semiconductor leads to charge redistribution,
it  does not change the total 
charge in it. Only if strain changes near the inhomogeneous region near the edge
of the depletion layer,  the total charge experiences the perturbation.

For comparison, we show the kernel function 
for a rough model of step-like dependence of $k_s$, where it is set to zero in the 
depletion region and to the value of the bulk semiconductor to the left of its edge, assumed to be 
infinitely sharp. The approximation allows analytical determination of  
$G_s$. As we see, for semiconductor this model is not very good, especially 
at room temperature where depletion region edge is not well-defined. However, it 
is good for the metal region since here any energetic perturbation is much less then 
the Fermi energy. As a result, for metal we can use the analytical expression 
for $G_m$, which is $G_m=k_m \exp (-k_m (z-z_i))$ for $z>z_i$. 
It is important to mention useful normalization conditions, which hold for any distribution of potential in the diode:
\begin{eqnarray}
\label{eq:kernel-normailzation}
\int_{-\infty}^{z_i} G_s (z) dz = 1, \\
\int_{z_i}^{\infty} G_m (z) dz = \xi_m \equiv \frac{1}{w_m} \left( \phi_{m1}(z_i) \phi'_{m2} (z_i)-\frac{\varepsilon_s}{\varepsilon_m}\phi_{m2} (z_i) \phi'_{m1}\right) \nonumber
\end{eqnarray} 

Let us discuss the in some details the deformation potential and piezoelectric
couplings. For semiconductor-contribution this is straightforward. The deformation coupling
describes the shift of the bottom of the conduction band minima. 
Its specific form depends on the 
crystal symmetry and the momentum position of the conduction band
\cite{Ivchenko-Pikus}. In any case, $V_{DP}$ is proportional to strain. Below, to be specific,
we will provide expressions for the case of GaAs with $z$-axis parallel
to its [111] crystallographic direction and longitudinal acoustic wave propagating along $z$. 
In this case 
\begin{equation}
\label{eq:DP111}
V_{DP}= \frac{E_1}{e} u_{zz},
\end{equation}
where $E_1$ is the deformation potential constant and
$u_{zz}$ is the only present component of strain. 

The piezoelectric potential is determined by the 
strain-induced piezoelectric polarization. For the mentioned geometry and acoustic wave polarization 
we obtain 
\begin{equation}
\label{eq:PZ111}
V_{PZ}= \frac{2 e_{14}}{\sqrt{3}\varepsilon \varepsilon_0} u_z,
\end{equation}
where $e_{14}$ is the piezoelectric constant of a cubic material and $u_z$ is the only present component of 
displacement in the considered longitudinal acoustic wave. 
As we see, the piezoelectric effect induces charge not only because of charge redistribution, but also due to 
direct induction of potential (the second term in the brackets of Eq.(\ref{eq:charge-expr})).
 
It is worth to mention that for the case of  
different crystallographic orientation, crystal symmetry or acoustic wave polarization the general structure of the expressions for 
deformation and piezoelectric potentials remains the same with the former proportional to strain and the latter proportional to 
displacement. Of course, in some cases some contributions vanish. For example, in GaAs there is no piezoelectric coupling 
for acoustic wave of any polarization propagating along [100] direction; deformation potential in this case is absent for transverse wave.

For the metal region consideration of the coupling of acoustic wave to electrons is more complicated than for semiconductor. This is 
because the deformation potential in a metal is considered as a perturbation of electron spectrum in some momentum point near the 
Fermi surface. Therefore, this value is, strictly speaking, momentum dependent. While considering screening, a momentum-averaged value is introduced to determine the charge perturbation
\cite{Gantmakher-Levinson,Abrikosov}. Its dependence on the strain components is determined by the symmetry of the metal Fermi surface. In fact, the corresponding 
constants are hardly known. This is because experiments on electron transport or ultrasound attenuation in metals  provide {\it screened} value of electron-phonon coupling averaged in a specific way
\cite{Abrikosov}. In the following, we will use in metal  
\begin{equation}
\label{eq:DP-metal}
V_{DP}= \frac{E_m}{e} u_{zz},
\end{equation}
keeping in mind that the effective constant $E_m$ has specific value dependent on the metal crystallographic orientation (for metal single crystals)
and acoustic wave polarization. By the order of magnitude, one can expect $E_m$ to be about few 
electronvolts.  

It is worth mentioning that in general we should not discard piezoelectric-like coupling in metal. 
It is usually done while considering electron scattering by phonons since efficient screening in metals cancels any macroscopic potential. However, the magnitude 
of the space charge induced under the screening does not vanish. In particular, this is seen from the expression for $G_m$, which provides finite value for the induced 
charge regardless of large value of $k_m$. 
In the following we do not include piezoelectric contribution in metal into consideration since no info is available of its presence and strength. However, one has to keep in mind that high-frequency acoustic wave detection by Schottky diode could reveal possible
piezoelectric-like coupling in metals. In principle, it can be distinguish from the deformation potential, since, similar to the semiconductor, the it should be proportional to the displacement rather than strain  

\section{Detection of the acoustic wave by the diode}

Naturally, the signal induced by an acoustic wave passing through the diode depends both on its intrinsic characteristics and the properties of the electrical circuit which includes the 
Schottky diode. We consider simple model circuit consisting of the diode and series resistance $R$ (see the insert of Fig.\ref{fig:1}).
Using Eq.(\ref{eq:charge-expr}) we can easily obtain equation for $\delta V$:
\begin{eqnarray}
\label{eq:circuit}
\frac{d \delta V}{dt} +\frac{\delta V}{RC} =\frac{dS}{dt} \\  
S=\left(  V_{PZ}(z_i) - \int_{-\infty}^{z_i} dz G_s (z) \left( V_{DP}(z) +V_{PZ} (z)\right) + 
\xi_m  V_{DP} ^{(m)}(z_i) \right), \nonumber
\end{eqnarray}
where the right hand side can be considered as a source caused by an acoustic wave, 
smallness of the screening length in the metal is taken into account, and superscript $(m)$ indicates the deformation potential in the metal. The   particular
form of the acoustic signal depends on the kind of the acoustic source. In high-frequency band the most popular one is a bipolar strain pulse generated with the use of picosecond ultrasonics technique 
\cite{ps-ultrasonics}. Alternatively, quasi-monochromatic acoustic waves can be produced by 
semiconductor superlattices illuminated by femtosecond laser pulses \cite{ps-ultrasonics} or 
sasers \cite{saser1,saser2,saser3}. Since in the linear response regime any acoustic signal can be 
presented as a plane wave superposition, in Eq.({\ref{eq:circuit}) we switch to the frequency domain and obtain 
\begin{equation}
\label{eq:circuit-freq}
\delta V_\omega = \frac{1}{1+i (\omega RC)^{-1}} S_\omega.
\end{equation} 
The intrinsic detection properties of the diode are reflected by the frequency dependence of
$S_\omega$. In fact, it is determined by the spatial broadening of the kernel function $G_s$. 
Assuming the plane-wave strain, we obtain 
\begin{equation}
\label{eq:S_omega}
S_\omega= - i \tilde{V}_{PZ} \left(1-J_s \exp(i\theta)\right) +\xi_m \tilde{V}_{DP}^{(m)} -
\tilde{V}_{DP}^{(s)}J_s \exp(i\theta),
\end{equation}
where $\tilde{V}_{PZ}$ and $\tilde{V}_{DP}^{(s,m)}$ are the amplitudes of the piezoelectric and deformation potentials (with the superscript labeling semiconductor and metal contributions). For the specific case of [111]-oriented semiconductor (Eqs.(\ref{eq:DP111},\ref{eq:PZ111},\ref{eq:DP-metal})) we have  
$\tilde{V}_{PZ}=2 e_{14} u_{zz}^{(0)}s \left(\sqrt{3} \varepsilon_s \varepsilon_0 \omega\right)^{-1}$ and
$\tilde{V}_{DP}^{(s,m)}= E_{s,m} u_{zz}^{(0)}/e$, where $s$ is sound velocity and $u_{zz}^{(0)}$
is the strain amplitude. In Eq.(\ref{eq:S_omega}) the overlap integral is introduced:
\begin{equation}
\label{eq:overlap}
J_s \exp (i\theta) =\int_{-\infty} ^{z_i} dz G_s(z) \exp (i\omega z/s).
\end{equation}
\begin{figure}
\centering
    \includegraphics[width=0.7\linewidth]{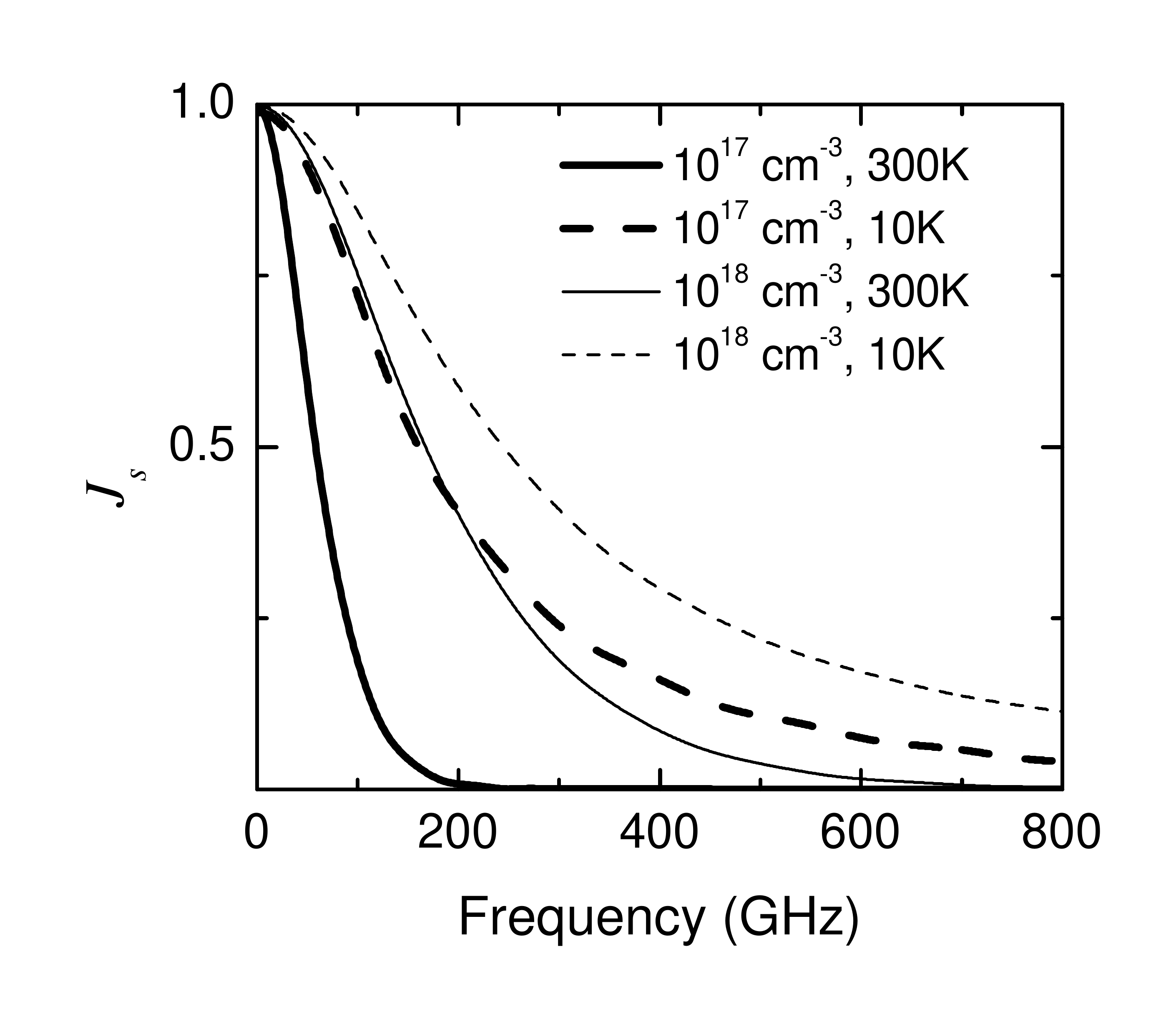}
\caption{The calculated overlap $J_s$ for various doping levels and temperature. 
}
\label{fig:overlap}
\end{figure}
The calculated frequency dependence of $J_s$ is shown in Fig.\ref{fig:overlap}. As it is expected, $J_s$ is
suppressed for frequencies corresponding to the acoustic wavelength smaller than the spatial localization length of the kernel $G_s$. The frequency dependence of $\theta$, which is not shown in a graph, reflects the phase  shift of the acoustic signal at the edge 
of the depletion layer and at the metal-semiconductor interface and corresponds roughly to $2\pi$ variation for 
frequency increase about $90$ and $26$~GHz for doping $10^{18}$ and $10^{17}$~cm$^{-3}$, respectively.    
\begin{figure}
\centering
    \includegraphics[width=0.7\linewidth]{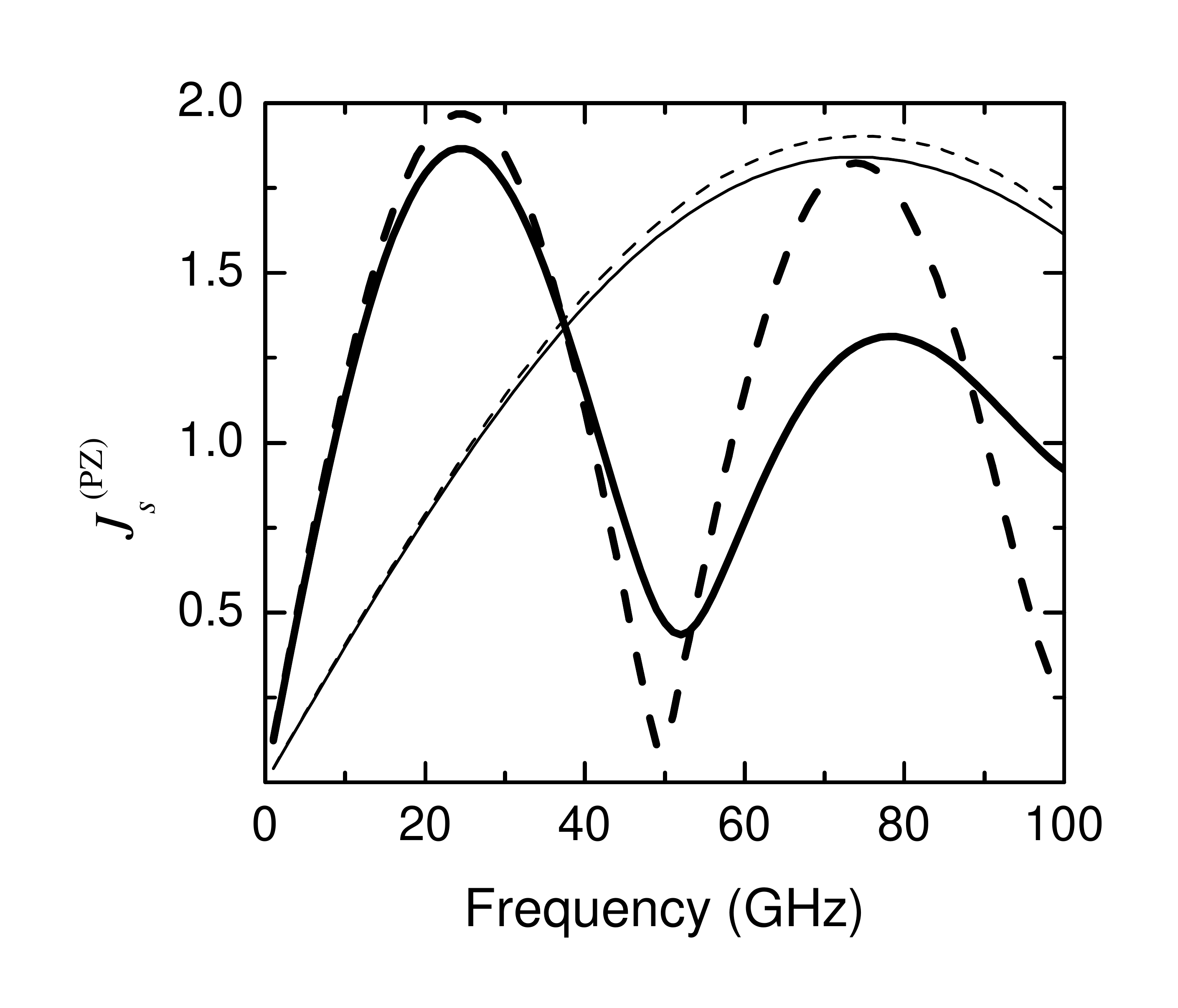}
\caption{The calculated value of $J_s^{(PZ)}$ for various doping levels and temperature.
The lines' legend is the same as in Fig.\protect\ref{fig:overlap}. 
}
\label{fig:overlap-pz}
\end{figure} 

If piezoelectric coupling is present in the structure, it commonly exceeds the deformation one for frequencies below a hundred gigahertz. 
For separate analysis of the piezoelectric contribution it is convenient to introduce the value $J_s^{(PZ)} \equiv 
|1 -J_s \exp (i \theta)|$. The frequency dependence of $J_s^{(PZ)}$ is shown in Fig.\ref{fig:overlap-pz}.   Naturally, it shows resonances corresponding to in-phase perturbation at the edge of the semiconductor depletion region and metal-semiconductor interface. Positions of these resonances can be easily predicted since the piezoelectric contribution to the diode response 
is determined by the parameters of semiconductor only,  which are usually well-known. 

It is worth to mention a special case of piezoelectric coupling and relatively low frequency acoustic wave, for which the acoustic wavelength is larger than both the broadening of $G_s$ and the thickness of the depletion layer. Here, $S$ becomes  proportional to strain. So, for the particular case of Eq.(\ref{eq:PZ111}) we have $S=
2 e_{14} u_{zz} L_{eff}(\sqrt{3}\varepsilon_0 \varepsilon_s)^{-1}$. If, in addition, if $(RC)^{-1}$ exceeds 
considerably the   characteristic acoustic frequency, $\delta V = S$. In other words, the electrical signal measures
directly the value of  strain in near-interface region.  For doping $10^{18}$~cm$^{-3}$, this approach can be valid for frequency up to several tens of gigahertz.

In diodes where piezoelectric coupling is absent, for example those employing non-piezoelectric semiconductors, like Si or Ge, or grown along certain crystallographic directions, like [001] GaAs, 
the situation is different. The resonances are expected in this case as well, but their location is difficult to predict
because of unknown value of the effective deformation potential constant in metal. 

For higher frequencies the deformation potential coupling is most efficient. In addition, as we see from Fig.\ref{fig:overlap}, 
the semiconductor contribution is suppressed for high frequencies. However, the metal contribution persists for any realistic frequency. This means that the actual frequency restrictions are set by the ability of high-frequency electronics   to measure the high-frequency electric signals. 


Summarizing the obtained results we can conclude that the acoustic wave detection by Schottky diodes can be described by a simple model where electrical response of the diode is 
caused by the displacement current induced by electrons screening the strain-induced perturbation. The actual upper frequency limit is set by the 
parameters of the current-registering equipment rather than internal diode properties due to the fast electronic response and small screening length in metal
contact of the diode. On the other hand, the semiconductor-side signal contributions are efficient, for common diode structures for frequencies below few
hundreds of gigahertz. These results will be an important guide for interpretation of the measured electrical diode response to an
acoustic perturbation as well as optimization of the Schottky diode acoustic wave detectors.

\begin{acknowledgments}

\end{acknowledgments}


%


\end{document}